\documentstyle[pre,aps]{revtex}

\newcommand{\be}{\begin{equation}}
\newcommand{\ee}{\end{equation}}
\newcommand{\om}{\omega}
\newcommand{\sgm}{\sigma}
\newcommand{\ra}{\rightarrow}
\newcommand{\prt}{\partial}

\begin{document}

\draft

\title{Self-similar approximations for trapped Bose-Einstein
condensate}
\author{V.I. Yukalov$^{1,2}$, E.P. Yukalova$^{1,3}$, and
V.S. Bagnato$^1$}

\address{$^1$Instituto de Fisica de S\~ao Carlos, Universidade de S\~ao Paulo \\
Caixa Postal 369, S\~ao Carlos, S\~ao Paulo 13560-970, Brazil}

\address{$^2$Bogolubov Laboratory of Theoretical Physics \\
Joint Institute for Nuclear Research, Dubna 141980, Russia}

\address{$^3$Department of Computational Physics,
Laboratory of Information Technologies \\
Joint Institute for Nuclear Research, Dubna 141980, Russia}

\maketitle

\begin{abstract}

An approximate solution to the Gross-Pitaevskii equation  for Bose-Einstein 
condensate in a spherical harmonic trap is suggested, which is valid in the 
whole interval of the coupling parameter, correctly interpolating between 
weak-coupling and strong-coupling limits. This solution is shown to be more 
accurate than the optimized Gaussian approximation as well as the 
Thomas-Fermi approximation. The derivation of the solution is based on the 
self-similar approximation theory. The possibility of obtaining interpolation
formulas in the case of nonspherical traps is discussed.

\end{abstract}

\vskip 1cm

\pacs{03.75.-b, 03.75.Fi}

\newpage

Bose-Einstein condensates of trapped atomic gases at low temperatures are 
described by the Gross-Pitaevskii equation (see reviews [1--3]). This 
nonlinear equation has no exact solution, and for practical applications 
one needs to resort to some approximations. The most commonly used such 
approximations are the Gaussian approximation and the Thomas-Fermi 
approximation. The former is asymptotically correct in the weak-coupling 
limit, while the latter, in the strong-coupling limit. Both these 
approximations are not accurate in the intermediate region of the coupling 
parameter. The Gaussian approximation can be improved by invoking an 
optimization procedure. Then its region of validity with respect to the 
coupling parameter essentially increases. However, it does not become exact 
in the limit of an infinite coupling. The Thomas-Fermi approximation, in 
addition to being incorrect at small and moderate coupling parameters, is 
always inadequate near the edge of an atomic cloud. There exist several 
suggestions [4--7] for correcting this approximation. But then the main 
advantage of the Thomas-Fermi approximation of being simple becomes lost. 
It would be desirable to have an approximation that would be valid for the 
whole range of the coupling parameters and would not require additional 
corrections.

The aim of the present paper is to derive an approximate solution to the
Gross-Pitaevskii equation with spherical symmetry, such that it would be
accurate for arbitrary values of the coupling parameter and also would
yield asymptotically exact solutions in both limits of weak as well as
strong coupling. We compare the derived solution with those obtained in
an optimized Gaussian approximation and the Thomas-Fermi approximation
and show that it is more accurate than the latter two. We also suggest
the ways of generalizing the method for nonspherical traps.

Let us consider a Bose-Einstein condensate of atoms with mass $m_0$ in a 
harmonic spherical trap of frequency $\om_0$. It is convenient to work with 
dimensionless quantities, measuring the radial variable
$r\equiv\sqrt{r_x^2+r_y^2+r_z^2}/l_0$ in units of the oscillator length 
$l_0\equiv\sqrt{\hbar/m_0\om_0}$. The related dimensionless ground-state wave
function $\psi_0(r)$ depends on the variable $r$. The dimensionless coupling 
parameter $g\equiv 4\pi a_sN/l_0$ is expressed through the ratio of the 
$s$-wave scattering length $a_s$ to the length $l_0$, with $N$ being the 
number of atoms in the trap. The radial wave function is defined as 
$\chi(r)\equiv\sqrt{4\pi}r\psi_0(r)$. In this notation, the stationary 
Gross-Pitaevskii equation, for a harmonic trap, reduces to the eigenvalue 
problem
\be
\label{1}
\hat H_r \chi =  E\chi \; , \qquad \hat H_r \equiv - \; \frac{1}{2}\; 
\frac{d^2}{dr^2} + \frac{r^2}{2} + \frac{g}{4\pi r^2}\; \chi^2 \; .
\ee
In order to be self-consistent, the solution to the problem (1) has
to satisfy the {\it self-consistency conditions} $E=(\chi,\hat H_r\chi)$
and $(\chi,\chi)=1$. To find an approximate solution to the problem (1), we 
shall employ the {\it self-similar approximation theory} [8--11], using the 
variant [12,13] designed for constructing crossover approximants satisfying 
the prescribed asymptotic conditions. Thus, we may notice that in the 
Hamiltonian there are two terms of different physical nature. The term $r^2/2$ 
corresponds to the harmonic trapping potential, while the last term is due to 
atomic interactions. The contribution of these two terms is different for 
different values of the variable $r$. There exists a crossover radius $r_c\equiv
(g/2\pi)^{1/4}$, separating the axis $r\geq 0$ onto two regions, where one of 
these terms plays the major role. For $r\ll r_c$, the interaction term prevails 
over the harmonic one, but for $r\gg r_c$, the main contribution comes from the 
harmonic term. It is straightforward to check that for small $r$ the solution to
Eq. (1) reads as an expansion $\chi(r)\simeq c_1 r+c_3 r^3+c_5 r^5$, as $r\ll 
r_c$, while for large $r$ it has the form $\chi(r)\simeq C r\exp(-r^2/2)$, as 
$r \gg r_c$. As is clear, the limit $r\ra 0$ is related to the strong-coupling 
limit $g\ra\infty$, when the interaction term dominates, while $r\ra\infty$ 
corresponds to the weak-coupling limit $g\ra 0$, where the harmonic term becomes
prevailing. Our aim is to find a solution that would be accurate for all $r$ 
and, respectively $g$, and would satisfy the prescribed asymptotic conditions.

The idea of the method, we shall use, is to construct a {\it self-similar
interpolation} between the asymptotically exact expressions. The name of the 
method comes from employing renormalization group in the form of group 
self-similarity [8--11]. All mathematical foundations of the theory and technical
prescriptions are expounded in detail in Refs. [8--13]. The result of the 
self-similar interpolation between the asymptotic expressions is the {\it 
self-similar approximant}
\be
\label{2}
\chi_*(r) = C r \exp\left ( -\; \frac{r^2}{2}\right ) \exp\left\{
ar^2 \exp(-br^2)\right \} \; .
\ee
Here $C$ is a normalization constant and the parameters $a$ and $b$ are defined 
by expanding Eq. (2) in powers of $r$ and substituting this expansion into Eq. 
(1). Equating the coefficients at the powers of the same order gives 
$a=1/2+(gC^2 -4\pi E)/12\pi,\; b=[2(1-2a)E-2(1-2a)^2 -1]/20a$. The energy $E=E_*$
in this approximation and the normalization constant $C$ are defined by the 
self-consistency conditions, that is from the equations $E_*=(\chi_*,\hat H_r
\chi_*)$ and $(\chi_*,\chi_*)=1$. By this construction, the self-similar 
approximant (2) possesses the correct behaviour at small $r\ll r_c$ and at large 
$r\gg r_c$, thus, interpolating between asymptotically exact expressions.

The accuracy of an approximation can be verified by calculating the {\it local 
residual} $R(r)\equiv(\hat H_r - E)\chi(r)$ and the {\it integral deviation}, 
$\sgm \equiv[\int_0^\infty |R(r)|^2dr]^{1/2}$, defining, respectively, the 
deviation of an approximate solution from the exact one at each given point 
$r$ and giving the integral measure of accuracy. Before calculating these, we 
shall recall some other known approximations in order to compare their accuracy 
with that of the self-similar approximation and also to explicitly demonstrate 
what difference in such a physical quantity as the atomic density $n(r)\equiv
\chi^2(r)/r^2$ results from the usage of different approximations.

When the coupling parameter $g$ is small, one may solve the nonlinear 
eigenproblem (1) by means of perturbation theory starting with the linear 
Hamiltonian. There exists an analytical continuation of linear modes to 
nonlinear stationary states [14]. Direct application of perturbation theory 
is valid only for asymptotically small coupling parameters $g\ra 0$. In order 
to make perturbation theory relevant for finite values of $g$, it is necessary 
to invoke an optimization procedure. Thus, one comes to optimized perturbation 
theory. This approach was formulated [15] for treating the systems whose 
particles strongly interact with each other. The theory has been applied to
various problems of quantum mechanics, statistical physics, condensed matter 
physics, and quantum field theory [3,11,15--19]. The optimization is realized 
by means of control functions [15]. In the present case, we may start from an 
approximate Hamiltonian with a harmonic potential $u^2r^2/2$ containing a trial 
parameter $u$. The ground-state eigenfunction of this Hamiltonian is of the 
Gaussian type, being
\be
\label{3}
\chi_G(r) = 2 \left ( \frac{u^3}{\pi}\right )^{1/4} r \exp\left ( -\;
\frac{u}{2}\; r^2 \right ) \; .
\ee
Then, applying some variant of perturbation theory, e.g. Rayleigh-Schr\"odinger 
theory, one can find a sequence $\{ E_k(g,u)\}$ of approximations for the energy.
For instance, the first approximation is 
\be
\label{4}
E_1(g,u) = \frac{3}{4}\left ( u + \frac{1}{u}\right ) + \frac{s}{2}\;
u^{3/2} \; , \qquad s \equiv \frac{2g}{(2\pi)^{3/2}}\; .
\ee
Control functions $u_k(g)$ are defined so that to render the sequence
$\{ E_k(g,u_k(g))\}$ convergent. For example, the optimization condition
$\prt E_k(g,u)/\prt u=0$ may be employed, resulting in the solution $u=u_k(g)$, 
which for the case (4) gives the equation $su^{5/2}+u^2-1=0$. If we stop at the 
first step of the optimized perturbation theory, then we get the optimized 
Gaussian approximation $E_G(g)\equiv E_1(g,u_1(g))$. Here we limit ourselves by 
this approximation, though higher-order corrections can also be obtained. 
Another popular approximate solution is the Thomas-Fermi approximation which is 
often used because of its simplicity. In that case, one neglects the kinetic 
term in the Gross-Pitaevskii equation (1), which yields $\chi_{TF}(r)=r
\Theta(r_0^2-r^2)\sqrt{(2\pi/g)(r_0^2-r^2)}$, where $\Theta(\cdot)$ is a 
unit-step function and $r_0\equiv 2E_{TF}$. The energy is obtained from the 
normalization condition for the function, which gives $E_{TF}=\frac{1}{2}
(15g/4\pi)^{2/5}$. Since the first of the self-consistency conditions is not 
satisfied, the Thomas-Fermi approximation is not self-consistent.

To compare the accuracy of the approximations described above, we calculate the 
local residual, integral deviation, and the related energies by using, 
respectively, the self-similar approximant (2), optimized Gaussian approximation
(3), and the Thomas-Fermi approximation $\chi_{TF}$. The calculations show that 
the self-similar approximant (2) possesses the lowest residual, providing the 
most accurate solution for the Gross-Pitaevskii equation. In the case of the 
coupling parameter $g=25$, the residual for the self-similar approximant gives 
$|R(r)|\leq 0.01$; for the optimized Gaussian approximation, $|R(r)|\leq 
0.4$; while for the Thomas-Fermi approximation, it diverges near the edge of the
atomic cloud. In the case $g=250$, for the self-similar approximant, we have 
$|R(r)|\leq 0.4$; for the optimized Gaussian approximation, $|R(r)|\leq 1.0$; 
and the residual for the Thomas-Fermi approximation diverges near the classical 
turning point. The calculation of the integral deviation confirms that the 
self-similar approximant has the best accuracy. The deviation $\sgm$ for this 
approximant does not exceed $2.02$ for all $g\geq 0$. It is very small, 
$\sgm\ll 1$, for the coupling $g$ from weak to moderate. It slightly increases 
with increasing $g$, reaching $\sgm=2.02$ at $g=2411$. Then it again decreases 
as $g$ continues to increase. For instance, $\sgm=1.04$ at $g=2500$. And again,
$\sgm\ll 1$ for $g\ra\infty$. For the optimized Gaussian  approximation, the 
deviation $\sgm$ slowly increases with rising $g$. Thus, $\sgm=3.63$ at $g=2500$.
Hence, this approximation works yet rather well even for sufficiently strong 
coupling parameters of the order $g\sim 10^3$. In the case of the Thomas-Fermi 
approximation, the deviation $\sgm$ is infinite.

Figure 1 presents the behaviour of the energy for the related approximants, 
with varying the coupling $g$. The optimized Gaussian approximation is 
asymptotically exact in the limit $g\ra 0$, while the Thomas-Fermi approximation 
becomes asymptotically exact for $g\ra\infty$. The self-similar approximation 
interpolates between these two limits, being the best approximation uniformly 
valid for all $g$. The optimized Gaussian approximation is quite reasonable
up to $g\sim 10^3$. The Thomas-Fermi approximation is bad for low $g$ and becomes
reasonable after $g\sim 100$. Figure 2 illustrates the atomic density in 
different approximations as a function of the variable $r$ for different 
couplings $g$. For $g=5$, the self-similar and optimized Gaussian approximations
are close to each other, being quite accurate solutions, which is confirmed 
by low residuals and small deviations $\sgm$. The Thomas-Fermi approximation
for such a relatively weak coupling is yet very inaccurate. For $g=25$, the 
self-similar and optimized Gaussian approximations are still close to each other,
and the Thomas-Fermi approximation starts approaching them. When $g=250$, all 
three approximations are of comparable accuracy, except that the Gaussian 
approximation slightly worsens near the trap center and the Thomas-Fermi 
approximation, at the edge of the atomic cloud. With varying the coupling from 
weak $g\ra 0$ to strong $g\ra\infty$, the self-similar approximation smoothly 
interpolates between the optimized Gaussian and Thomas-Fermi approximations, 
remaining the best approximation for all $g$, which is practically 
indistinguishable from the exact numerical solution of Eq. (1). A special 
caution is in order when the studied functions are small, as it happens for the 
asymptotic tails of our solution. Therefore, it is necessary to pay a particular
attention to such asymptotic tails, checking if the {\it relative} difference is
asymptotically small. Fortunately, for our case, the relative deviation of the 
self-similar approximation $\chi_*(r)$ from the exact numerical solution tends 
to zero at large $r$ as $\chi_*(r)/\chi(r)-1\simeq ar^2\exp(-br^2)\ra 0$,
that is, the approximate and exact solutions asymptotically coincide.

It is also worth commenting on the possibility of using the self-similar
approximation technique for nonspherical traps. There are two ways of applying 
this method to equations containing more than one variable. One way is by 
reducing the equation in several variables to an effective equation of one 
variable, e.g. by means of an averaging procedure [3,14]. Another possibility 
could be by looking for an approximate solution to the three-dimensional 
Gross-Pitaevskii equation by constructing a trial function that is factorized 
with respect to its variables, as is done for an anisotropic trap in Ref. [20]. 
It is also possible to invoke trial functions at the {\it intermediate} step for
deriving {\it analytical} expressions for energies, which would interpolate 
between weak-coupling and strong-coupling limits. For this purpose, one should 
derive asymptotic expansions for the energy in the regions of small and large 
coupling parameters, and then construct self-similar approximations 
interpolating between these asymptotic expansions. As an example, we present the
results of such an interpolation procedure for the ground-state energy of atoms 
in a {\it cylindrical} trap with the aspect ratio $\nu\equiv \om_z/\om_r$, 
where $\om_z$ is the longitudinal trap frequency and $\om_r$ is its transverse 
frequency. We give here the first three self-similar interpolative approximants 
for the energy measured in units of $\om_r$,
$$
E_1^* =a_0\left ( 1 + AG\right )^{2/5} \; , \qquad E_2^* = a_0 \left [
\left ( 1 + A_1 G\right )^{6/5} + A_2 G^2 \right ]^{1/5} \; ,
$$
\be
\label{5}
E_3^* = a_0 \left\{ \left [ \left ( 1 + B_1G\right )^{6/5} +
B_2 G^2 \right ]^{11/10} + B_3 G^3 \right\}^{2/15} \; ,
\ee
where $G\equiv 2g\nu/(2\pi)^{3/2}$, $g\equiv 4\pi a_s N/l_r\;(l_r\equiv
\sqrt{\hbar/m_0\om_r})$, and
$$
Aa_0^{5/2} = 1.746928 \; , \qquad A_1a_0^{25/6}=2.533913(2+\nu^2)^{5/6} \; , 
\qquad A_2a_0^5=3.051758 \; ,
$$
$$
B_1a_0^{125/22}\left ( 2+ \nu^2\right )^{5/66} = 1.405455\left ( 8 +
12\nu^2 + \nu^4\right )^{5/6} \; , \qquad
B_2a_0^{75/11} = 6.619620 \left ( 2 +\nu^2\right )^{10/11} \; ,
$$
$$
B_3a_0^{15/2} = 5.331202 \; , \qquad a_0 = 1 + \nu/2\; .
$$
Although the derivation of such formulas requires some work, but after being 
derived, they can strongly facilitate the overall consideration, since the 
energy is now given in sufficiently simply {\it analytical} form that is 
convenient to study with respet to the dependence on the coupling parameter 
$g$ and the trap aspect ratio $\nu$. We have estimated the accuracy of these 
formulas in the intervals $0<g<10000$ and $0<\nu<100$. The maximal percentage 
errors are between $4-12\%$ for $E_1^*$; between $2-5\%$ for $E_2^*$; and of 
order $1\%$ for $E_3^*$. 

In conclusion, by employing the self-similar approximation theory [8--13], we 
have found an approximate solution of the Gross-Pitaevskii equation for a 
spherical trap. This solution, presented by the self-similar approximant (2), is
compared with the optimized Gaussian approximation and Thomas-Fermi approximation.
It is shown that among these three approximations the self-similar approximant 
provides the best accuracy for all couplings $g$, tending in the limits of weak 
and strong couplings to the corresponding asymptotically exact solutions.

\newpage

\begin{center}
{\large{\bf Figure Captions}}
\end{center}

\vskip 1cm

{\bf Fig. 1}. Condensate energy (in dimensionless units) as a function
of the coupling parameter for the self-similar approximation (solid
line), optimized Gaussian approximation (dotted line), and for the
Thomas-Fermi approximations (dashed line).

\vskip 1cm

{\bf Fig. 2}. Atomic density (in dimensionless units) as a function
of the radial (dimensionless) variable, corresponding to the
self-similar approximant (solid line), optimized Gaussian approximation
(dotted line), and to the Thomas-Fermi approximation (dashed line) for
different coupling parameters: (a) $g=5$; (b) $g=25$; (c) $g=250$.

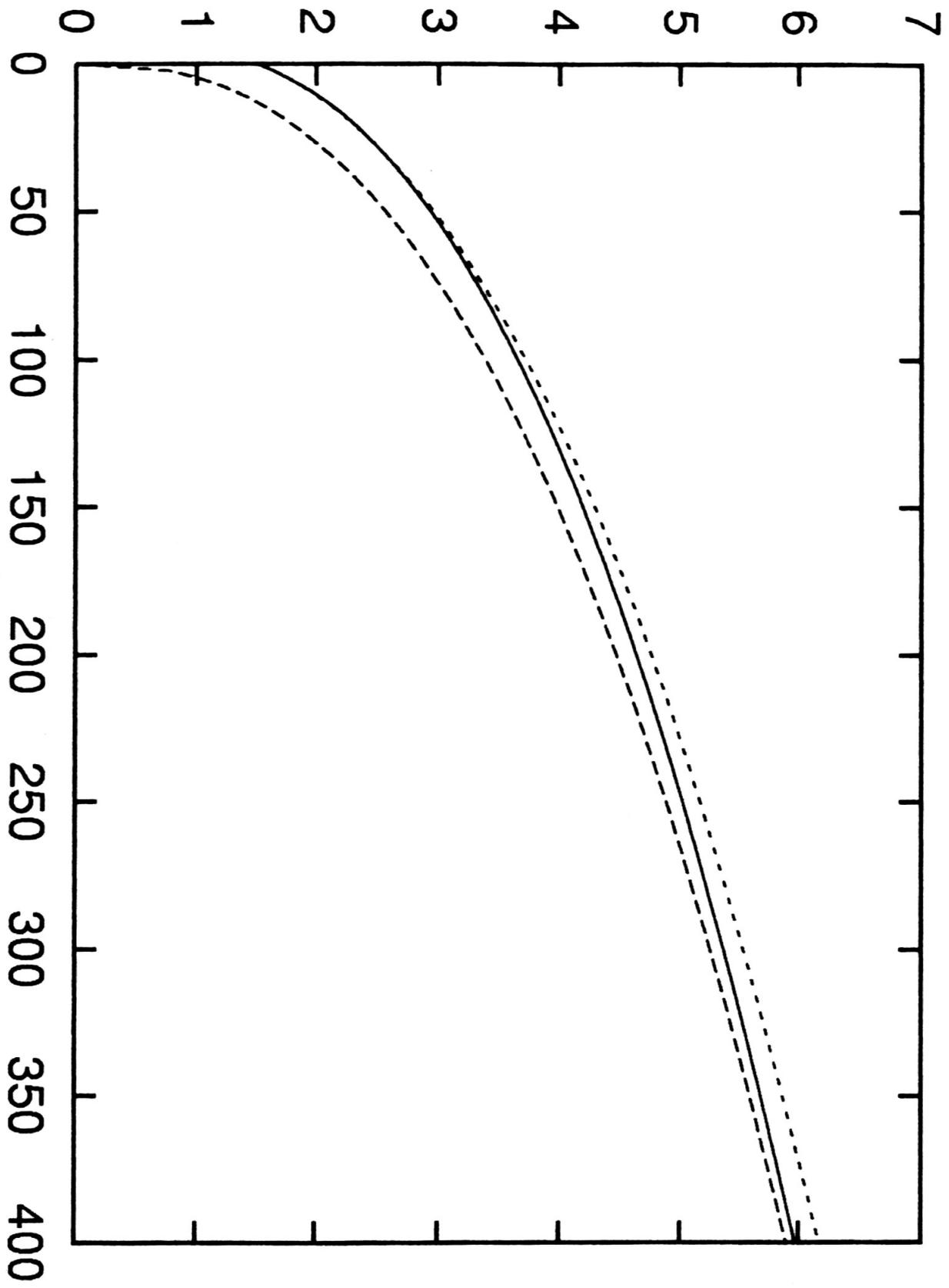
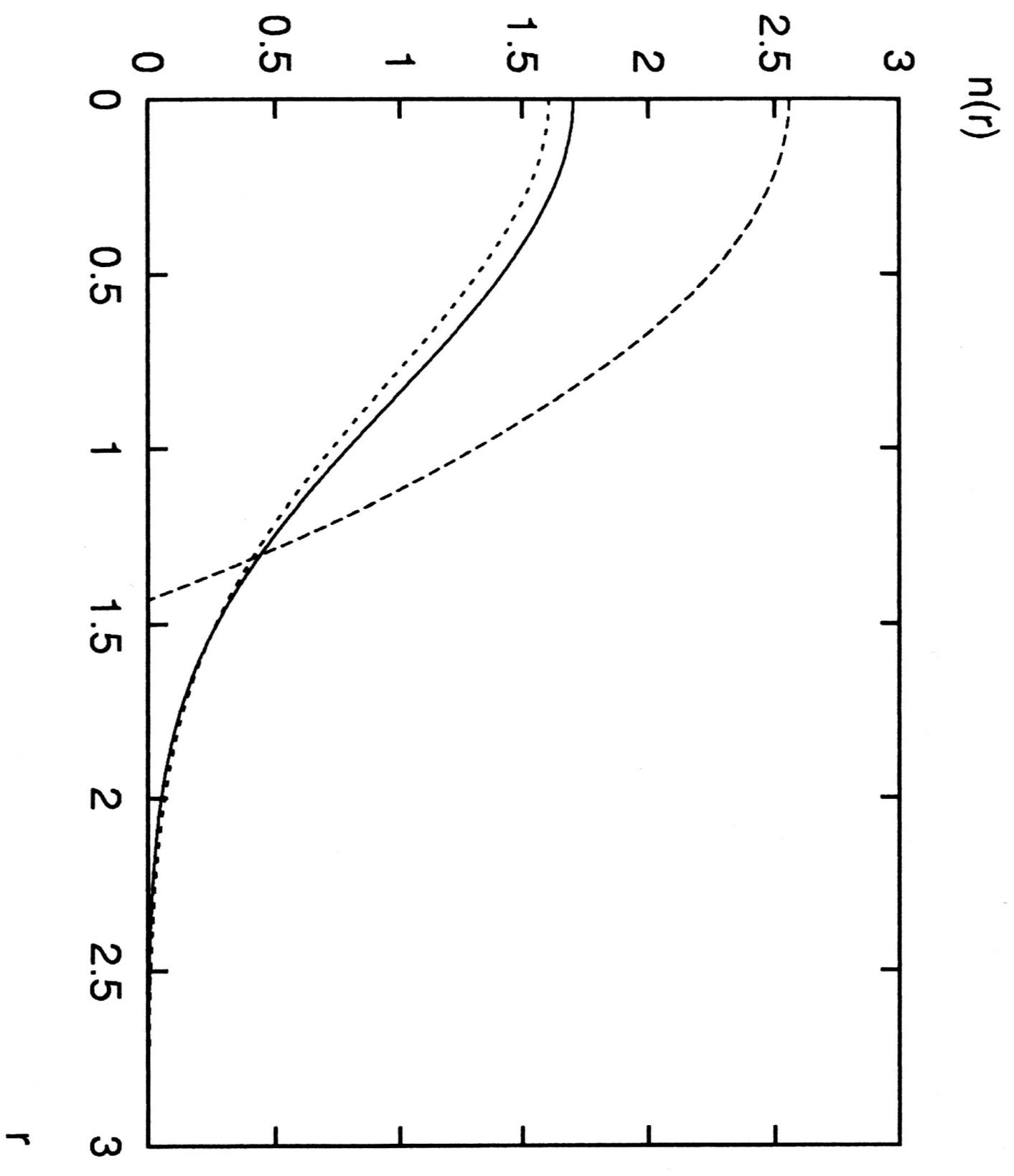
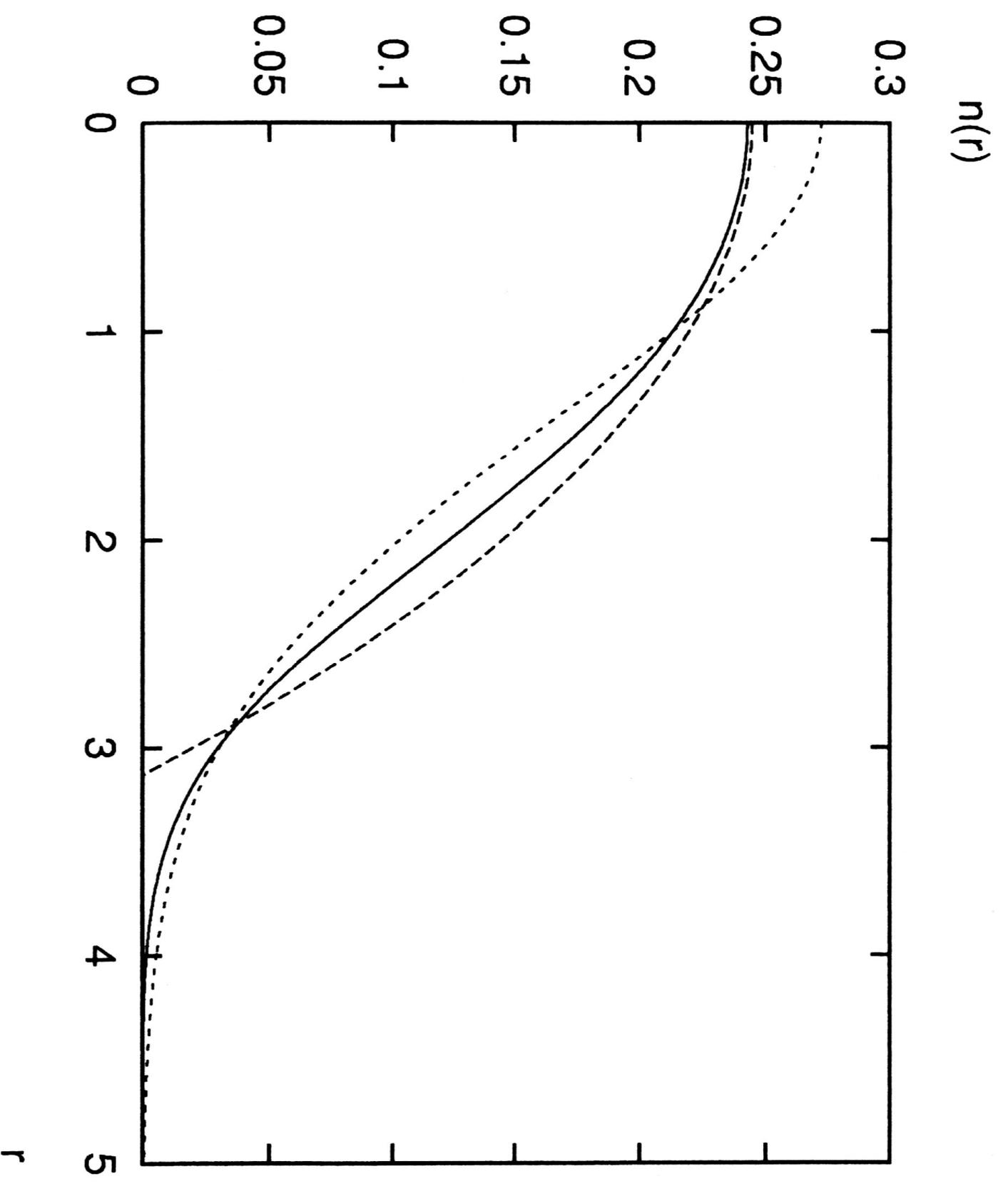
\end{document}